%% file: ms.tex
\documentclass[apjl]{emulateapj}
\usepackage{apjfonts}
\usepackage{epsfig}

\bibpunct{(}{)}{;}{a}{}{,}

\begin{document}

\title{IRS 16SW - A New Comoving Group of Young Stars 
in the Central Parsec of the Milky Way}

\author{
J. R. Lu,
A. M. Ghez\altaffilmark{1},
S. D. Hornstein,
M. Morris,
E. E. Becklin
}

\affil{UCLA Department of Physics and Astronomy, Los Angeles, CA 90095-1562}
\email{jlu, ghez, seth, morris, becklin@astro.ucla.edu}
\altaffiltext{1}{UCLA Institute of Geophysics and Planetary Physics, 
Los Angeles, CA 90095-1565}

\begin{abstract}
One of the most perplexing problems associated with the
supermassive black hole at the center of our Galaxy is
the origin of the young stars in its close vicinity.
Using proper motion measurements and stellar number density
counts based on 9 years of diffraction-limited K(2.2 \micron)-band
speckle imaging at the W. M. Keck 10-meter telescopes, 
we have identified a new comoving group of stars, which we call the
IRS 16SW comoving group, located 1\farcs9 (0.08 pc, in projection)
from the central black hole.
Four of the five members of this comoving group
have been spectroscopically identified as massive young stars,
specifically He I emission-line stars and OBN stars.
This is the second young comoving 
group within the central parsec of the Milky Way to be recognized and is the 
closest, by a factor of 2, in projection to the 
central black hole. These comoving groups
may be the surviving cores of massive infalling star clusters
that are undergoing disruption in the strong tidal field
of the central supermassive black hole. 
\end{abstract}


\section{Introduction}

The central parsec of our Galaxy harbors not only a supermassive
black hole (SBH) of mass 
$\sim$3.7 $\times$ 10$^6$ M$_{\sun}$ ({Ghez} {et~al.} 2003, 2005; 
{Sch{\" o}del} {et~al.} 2003), 
but also a cluster of young, massive stars within the
sphere of influence of this black hole. The young stars
include $\sim$40 He I emission-line stars 
identified from spectra as blue supergiants (Of), 
luminous blue variables (LBVs),
and Wolf-Rayet (WN/C) stars with masses ranging from 30 to 120 M$_{\sun}$, 
ages of 2-7 Myr, and distances limited to 1-10\arcsec ~from the SBH 
(e.g. {Allen}, {Hyland}, \& {Hillier} 1990; {Paumard} {et~al.} 2004).
Dynamical studies of these He I emission-line stars 
used the three-dimensional velocities to show that several, including
the brightest members of the IRS 16 complex such as 16C and 16SW,
exhibit a coherent, clockwise rotation pattern
({Genzel} {et~al.} 2000)
and may all lie on a common orbital plane highly inclined to our line
of sight ({Levin} \& {Beloborodov} 2003).
In addition to this clockwise plane, 
the counter-clockwise orbiting He I emission-line stars
are consistent with a second plane oriented nearly face-on to our
line of sight ({Genzel} {et~al.} 2003).
Within the counter-clockwise plane, IRS 13 has been 
identified as a comoving, compact cluster of massive young stars
located $\sim$4\arcsec ~from the SBH 
({Maillard} {et~al.} 2004). 

The origin of these young stars is puzzling given that the 
local gas densities are orders of magnitude too low to overcome the
tidal shear of the SBH and collapse to form stars 
({Sanders} 1992; {Morris} 1993). 
Several possible solutions have been proposed
including scenarios that enhance the local gas densities such as 
a self-gravitating accretion disk 
({Levin} \& {Beloborodov} 2003)
or infalling and colliding dense gas clouds 
({Genzel} {et~al.} 2003),
both of which allow stars to form {\it in situ}. 
However, no such sufficiently dense gas clouds have
been observed. Alternatively, stars might form at larger radii where
the tidal forces are less extreme and
migrate inwards through dynamical friction as part of a
massive stellar cluster 
({Gerhard} 2001; {Kim} \& 
{Morris} 2003; {Portegies Zwart}, {McMillan}, \&  {Gerhard} 2003;
{McMillan}, \&  {Portegies Zwart} 2003). 
It has been suggested that for this cluster to migrate
to within a few arcseconds of the SBH within the lifetime of the massive
young stars observed, the stellar cluster would need to be so dense as
to undergo core collapse and contain an intermediate mass black hole 
(IMBH; {Hansen} \& {Milosavljevi{\' c}} 2003; 
{Kim}, {Figer}, \& {Morris} 2004; {G{\" u}rkan} \& {Rasio} 2005). 
Although no clear cut case for an IMBH 
has yet emerged, the IRS 13 cluster is  coincident
with a bright, discrete X-ray source 
({Baganoff} {et~al.} 2003; 
{Muno} {et~al.} 2003) and 
may be the remaining core of an infalling cluster harboring such an
IMBH ({Maillard} {et~al.} 2004). 

The orbital kinematics of the young stars provides
important insight into their origins.
In this paper we report the discovery of a new comoving group of stars
within the clockwise plane
which includes the bright young He I emission-line star, IRS 16SW. 
Diffraction-limited observations 
and data reduction are described in \S\ref{obsdata}. Section \ref{results}
reports the discovery and quantitative significance of the IRS 16SW
comoving group. Finally, \S\ref{discussion} addresses the implications 
of this result
for the proposed theories for the origins of young stars at the Galactic
Center and discusses the possibility
that this group is the remnant core of a infalling cluster.

\section{Observations \& Data Analysis} \label{obsdata}

Speckle imaging observations of the Galaxy's central stellar cluster were 
taken in the K-band ($\lambda_{o}$=2.2 \micron, 
$\Delta\lambda$=0.4 \micron) using the facility near-infrared
camera, NIRC ({Matthews} {et~al.} 1996), on the Keck I 10-meter telescope.
Data sets taken between 1995 June and 2003 September, detailed in 
{Ghez} {et~al.} (1998, 2000, 2005), and new 
data sets taken on 2004.34, 2004.56, and 2004.66, consist of $\sim$10,000
short (t$_{exp}$ = 0.1 sec) exposures per observing run with a plate
scale of 20.40 $\pm$ 0.04 mas/pixel and a 
5\farcs22 $\times$ 5\farcs22
field of view. Frames are combined using a 
weighted shift-and-add technique (Hornstein et~al. in prep)
to form a final high resolution map for each data set 
(see {Ghez{ {et~al.} 2005 for details).
The data from each run were 
also divided into three sub-sets to create ``sub-maps'' used to determine
positional and brightness uncertainties.

\begin{figure*}
\epsscale{1.0}
\plotone{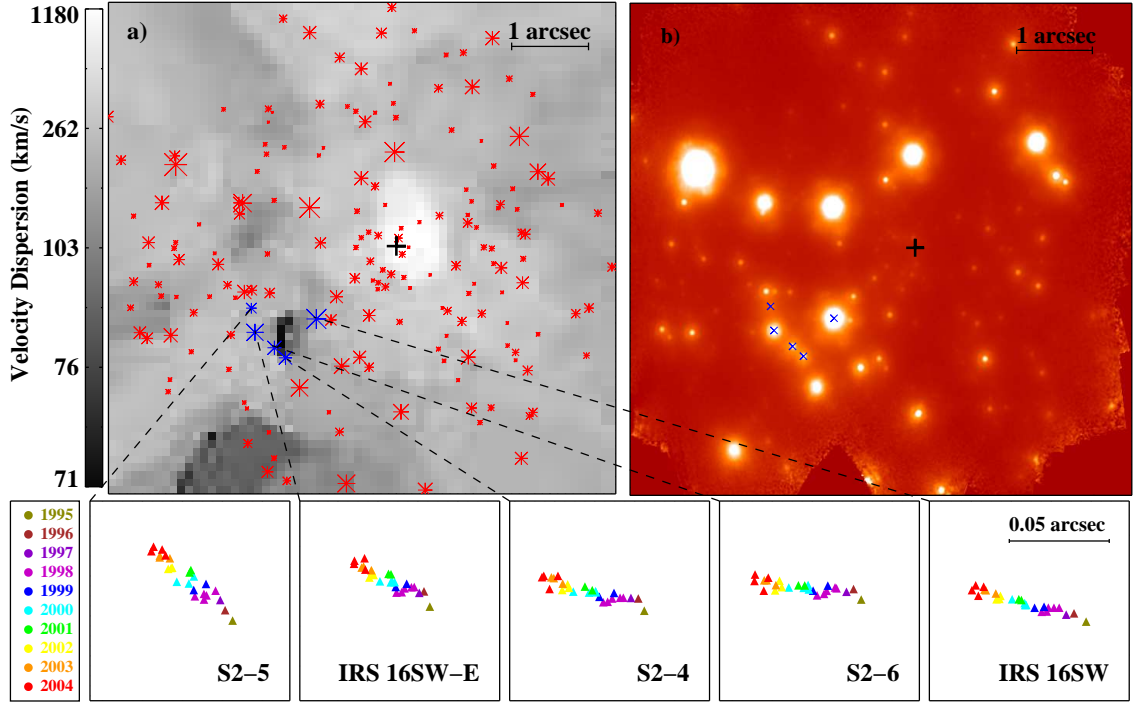}
\caption{
{\it (a):} The positions of all stars in our sample (asterisks)
overlaid on a map of the two-dimensional velocity dispersion
(grayscale). The sizes of the asterisks represent the stars' 
2.2\micron ~brightness. The black region is a minimum in the
velocity dispersion, located $\sim$2\arcsec ~from Sgr A* (black cross),
and is caused by 5 comoving stars (blue), which define the newly
identified IRS 16SW comoving group.
{\it (b):} A K[2.2\micron]-band speckle image showing the clustering
of bright sources at the position of the IRS 16SW comoving group. 
Group members are marked with blue Xs.
{\it (Panels):} The proper motions of the IRS 16SW comoving group members.
In each  0\farcs1 $\times$ 0\farcs1 panel, the stellar positions
are plotted with different years of data labeled with different colors.
}
\label{fig:vdisp_image}
\end{figure*}

Sources were identified using
the IDL PSF fitting routine, {\it StarFinder} ({Diolaiti} {et~al.} 2000). 
{\it StarFinder} generates a PSF from several bright stars in the field
and cross-correlates the resulting PSF with the image. Candidate 
sources are those with correlation peaks above 0.7 in the main maps and 
above 0.5 in the sub-maps. Only sources detected in all 3 submaps are
included in the final source list for each observation. The coordinate
system for each list is transformed to a common local reference frame
by minimizing the net offsets of all stars as described in 
{Ghez} {et~al.} (1998, 2005). Centroiding uncertainties are 
$\sim$1 mas, while alignment uncertainties range from
$\sim$1-5 mas. The final relative positional
uncertainty is the quadrature sum of the centroiding and alignment
uncertainties and is $\sim$2 mas for the bright (K$\lesssim$13.5)
stars near IRS 16SW. 
Proper motions are derived by fitting lines to the 
positions as a function of time, weighted by the positional 
uncertainties. We conservatively require that only sources detected
in 9 or more epochs, out of 22 total epochs,
are included in the final sample. 
This results in a final sample of 180 stars, which have an 
average total proper motion uncertainty of 0.53 mas/yr
for all sources located beyond one arcsecond of the central SBH.
All proper motions were converted to linear velocities using a 
distance of 8 kpc and the uncertainty in this distance is not included
in the velocity uncertainties ({Reid} 1993).

\section{Results} \label{results}

A two-dimensional velocity dispersion map of the stars in the sample
reveals a minimum located between IRS 16SW and IRS 16SW-E 
(Fig.~\ref{fig:vdisp_image}a; grayscale). The velocity dispersion map
is produced by calculating, at each position separated by 0\farcs1, 
the following quantity for the nearest 6 stars: 
$\sigma^2_{intrinsic}$ = $\sigma^2_{measured}$
- $\sum_{i=0}^N$ [error$^2$(v$_{x,i}$) + error$^2$(v$_{y,i}$)] / [2*(N-1)],
where the first term is the dispersion of the measured proper motions
and the second term removes the bias introduced by the uncertainties
in the proper motion measurements. The minimum in the velocity dispersion
map is insensitive to the number of stars used in the calculation;
using the nearest 5 to the nearest 8 stars produces a similar result. 
The significance of the velocity dispersion minimum is determined by 
comparing it to the velocity dispersion of stars in the sample that are
at comparable radii (1\arcsec $\le$ r$_{2D} \le $2\farcs6). Because
the young stars are known to show some level of dynamical
anisotropy due to coherent
rotation about the SBH ({Genzel} {et~al.} 2003),
we restrict the comparison sample to known late-type stars 
({Figer} {et~al.} 2003; {Ott} 2003). 
The minimum in the velocity dispersion is significantly lower 
(4.6$\sigma$) than the field velocity dispersion.

The velocity dispersion minimum arises from a comoving group of stars;
to formally define the members of the comoving group, we must
first eliminate those stars which appear near the group due to
projection effects. Formal membership is determined by
considering the difference between the velocity of each individual star
and the group's average velocity. Within the region of the velocity
dispersion minimum, only 5 stars have velocity offsets
that are consistently $\le 2\sigma$. Using only these 5 stars
to re-define the values of group velocity and velocity dispersion,
we find no additional stars with a total velocity offset less
than 3.5$\sigma$ within a 1\farcs1 search radius.
We therefore define these 5 stars, which include IRS 16SW, as the 
members of the comoving group 
(Table \ref{tab:members}, Fig.~\ref{fig:vdisp_image}-Panels).
The IRS 16SW comoving group has an average and RMS distance
from Sgr A* of 1\farcs92 and 0\farcs43, respectively, 
and a velocity dispersion of 
36 $\pm$ 13 km/s in RA and 38 $\pm$ 13 km/s in DEC.

\input{tab1.tex}

There are two additional, independent lines of evidence supporting
the existence of this comoving group. First, two of the 
group members, IRS 16SW and IRS 16SW-E, have identical
radial velocities ({Ott} 2003); the other members, unfortunately,
have no measured radial velocities. Second, the stellar number density
counts show an enhancement at the position of the IRS 16SW comoving group
(see Fig.~\ref{fig:vdisp_image}b). Since the stellar number counts are
strongly affected by the varying sensitivity across the field from the
halos of bright stars, we consider only stars with 
K$<$13.5, corresponding to 99$\%$ completeness, for this analysis.
Using an aperture of 0\farcs55, which is the minimum radius necessary
to encompass the IRS 16SW comoving group, 
we generated a stellar number density map.
Positions including the IRS 16SW comoving group members
show the highest stellar number density, which corresponds to  
6 sources per aperture. The probability of such an enhancement at this
radius over the field's average bright star density, 
2 sources per aperture, is relatively small (0.03).

Members of the IRS 16SW comoving group appear to be young, massive stars.
The two brightest members have spectra that classify them as
He I emission-line stars identified as massive, young 
stars of type Ofpe/LBV or WR ({Krabbe} {et~al.} 1995). 
An additional two stars, S2-6 and S2-4, have recently been 
proposed, from spectroscopic observations, as
OBN stars, which are more luminous, nitrogen-rich OB stars 
({Paumard} {et~al.} 2004).
The identification of four of the five members as young, massive stars
further supports the existence of this comoving group.
The remaining member, S2-5, lacks any spectral measurements to 
definitively classify it; 
however, its absolute K-magnitude, assuming A$_v$ = 30, 
is consistent with either a late-type giant
(M3 or M4 III) or a young massive star (WR or OB). 
The latter possibility is favored by 
narrow-band photometry through a CO feature, 
which shows no evidence of the CO absorption characteristic 
of late-type giants (Ott 2003). Furthermore, S2-5's dynamical association
with other young stars suggests that it is also a young star.

A model-dependent estimate of the IRS 16SW comoving group's three-dimensional 
distance from the SBH suggests that the line-of-sight component is
negligible.  
Specifically, IRS 16SW and IRS 16SW-E are supposed members of
the clockwise plane identified by 
{Levin} \& {Beloborodov} (2003) and {Genzel} {et~al.} (2003).
If we assume that all the members of the comoving group lie on this plane,
then their line-of-sight offsets from Sgr A* range from
-1\farcs1 to +0\farcs5, with an average values of -0\farcs4 (-0.016 pc).
This is small compared
to their plane-of-the-sky offsets from Sgr A* 
and places the group along the line of nodes where the
plane of the sky intersects the orbital plane.  Therefore, in this 
model, the physical distance between the IRS 16SW comoving group and 
the SBH is $\sim$0.08 pc.

It is highly unlikely that an inclined plane of stars orbiting the 
SBH could give rise to apparent comoving groups at the line of
nodes.  To quantify the likelihood that observational biases
might lead to apparent coherence,
Monte Carlo simulations were performed of 1000 stars with the 
following assumptions: 1) stars 
are distributed with uniform surface density in the plane out to
a radius of 7\arcsec, which is the outer edge imposed by
{Genzel} {et~al.} (2003), and 2) the stars are in circular orbits. 
Results from the simulation show that the line of nodes,
where the IRS 16SW comoving group lies,
is a point of high velocity dispersion, not low velocity dispersion as 
is observed. Given the above assumptions, the probability of finding
an apparent comoving group by random chance 
projected along the line of nodes within 4\arcsec ~of the SBH
is less than 1 in 10$^4$. This suggests that the observed IRS 16SW 
comoving group members are dynamically associated with each other 
beyond simply belonging to the clockwise orbital plane.

\section{Discussion \& Conclusions} \label{discussion}

The IRS 16SW comoving group is now 
the second such grouping found within the central parsec of the 
SBH. The first, IRS 13, has been proposed as a compact,
massive star cluster located 3\farcs6 in projection from Sgr A*
and possibly harboring an IMBH
({Maillard} {et~al.} 2004; {Eckart} {et~al.} 2004). 
Although the IRS 16SW comoving group is assigned to the clockwise plane and
the IRS 13 cluster is assigned to the counter-clockwise plane, 
both contain young, massive stars, 
indicating that the two groups have comparable ages,
strongly suggesting a similar formation mechanism.
However, there are distinct differences between the two comoving groups.
Unlike the IRS 13 cluster, which
has a high stellar number density of $\sim$40-80 objects/arcsec$^{2}$
({Eckart} {et~al.} 2004),
the IRS 16SW comoving group shows only a slight stellar number density
enhancement ($\sim$7 objects/arcsec$^2$ to a similar 
limiting magnitude of $\sim$16.5). Likewise, while the IRS 13 
complex sits on a bright complex of dust, ionized gas, and possibly an
unresolved population of stars, the IRS 16SW comoving group shows
no equivalently bright halo in the K-band. 
Any formation scenario
for the young, massive stars in the galactic center must explain
the presence of both groups.

One proposed {\it in situ} formation mechanism invokes self-gravitating
accretion disks present a few million years ago but since consumed 
by star formation, accreted onto the SBH, or dissipated via 
winds from the young, massive stars 
({Morris} 1993; {Levin} \& {Beloborodov} 2003). The disk
would produce stars in circular orbits in a common orbital plane.
The low velocity dispersion of the IRS 16SW comoving group along
the line of nodes suggests either a non-uniform distribution within
the plane, which would indicate that star formation was clumpy,  
or non-circular orbits, which would argue against 
accretion disk formation scenarios.
It remains to be determined whether two accretion disks can assemble,
form massive clusters of stars, and dissipate within only a few million 
years of each other. 
Another {\it in situ} formation
scenario, colliding gas clouds, cannot be supported or disputed
observationally as no detailed simulations of star formation during
such interactions have been performed. 

Alternative scenarios invoke the formation of massive star clusters 
at larger radii which spiral in via dynamical friction.
As a result of mass segregation and possible core
collapse, the surviving core would contain primarily massive stars.
If the IRS 16SW comoving group is a bound cluster, then the
virial, Bahcall-Tremaine ({Bahcall} \& {Tremaine} 1981), and 
Leonard-Merritt ({Leonard} \& {Merritt} 1989) 
projected mass estimators yield a cluster mass of 
$M\sim10^4 M_{\sun}$, 
which is an upper limit if the cluster is not bound. 
Similarly, for the simple assumption that the cluster acts as a point mass,
observations of the reflex motion of 
Sgr A* rule out clusters of $>10^4 M_{\sun}$ at the projected distance
of the IRS 16SW comoving group, assuming circular orbits 
({Reid} \& {Brunthaler} 2004). 
In order for the IRS 16SW comoving group
to remain bound, the required mass density to overcome
the tidal forces at 0.08 pc from Sgr A* is 
$\sim4\times10^9 M_{\sun}/pc^3$ yielding a cluster mass 
lower limit of $M_{lower}\sim10^5 M_{\sun}$ assuming
a uniform density cluster for an order of magnitude calculation. 
Therefore, if the comoving group is indeed situated only 0.08 pc
from Sgr A*, then it is most likely unbound.

If the IRS 16SW comoving group is not a bound cluster, then it is 
difficult to understand how such a comoving group can have survived
in the extreme tidal field of the supermassive black hole.
One possibility is that the group is the dissipating remnant core
of a tidally disrupted cluster.
{Gerhard} (2001) shows that,
for the remaining unbound core of a cluster with 
extent $\alpha$r$_t$, where r$_t$ is the tidal
radius, the timescale for orbital phases to spread by $\sim\pi$
is of order $t_{spread} \sim \frac{t_{orbit}}{\alpha}$. 
The circular orbital period at
the group's distance from Sgr A* is $<$ 1000 years, suggesting that
any clustering should be disrupted within a few thousand years.
If the IRS 16SW comoving group is the 
remaining core of a massive star cluster, then the disruption may
have begun at larger radii ($\sim0.4$ pc), where the furthest
young He I emission-line stars are observed, while the core continued to
migrate in via dynamical friction ({Gerhard} 2001). Given the range of 
orbital timescales, $\sim$10$^3$-10$^4$ yrs for 0.04-0.4 pc, 
and the present existence of a comoving
group, disruption would have occurred only a few tens of thousands
of years ago.  The other young, massive stars that seem to lie
on the same orbital plane as the IRS 16SW comoving group may be other
members of the disrupted cluster core which are spreading in phase. 
IRS 13 and the associated stars in the 
second orbital plane would then be a second infalling cluster in an
earlier stage of disruption. 
Increasing the number of known young stars and measuring the
spread in phases would provide a better estimate of the time since
disruption. 
Simulations of the final disruption of 
an infalling cluster are needed to fully test this possibility,
and to determine if such a disruption can also give rise to the
cusp of OB stars within 0\farcs5 of Sgr A* (e.g. {Ghez} {et~al.} 2003,
{Eisenhauer} {et~al.} 2005).
The short duration of the final cluster disruption, our
observations at such a unique epoch, and the presence of two young
comoving groups on different orbital planes suggest that either the 
frequency of formation and infall of massive clusters 
is high ({Portegies Zwart} {et~al.} 2002) 
or, more likely, there was an epoch of triggered star 
formation several million years ago ({Figer} {et~al.} 2004) 
which produced two massive clusters at similar distances from
the central supermassive black hole. 
In conclusion, the IRS 16SW comoving group provides further evidence
for formation scenarios that involve clustering whether from infalling 
massive star clusters or {\it in situ} formation. 

\acknowledgements

Support for this work was provided by NSF grant AST-0406816 and
the NSF Science \& Technology Center for AO, managed by UCSC
(AST-9876783) and the Packard Foundation. 
The W.M.~Keck Observatory is operated as a scientific 
partnership among the California Institute of Technology, the University 
of California and the National Aeronautics and Space Administration. 
The Observatory was made possible by the generous financial support of 
the W.M.~Keck Foundation. 

\bibliography{}

\end{document}

%% file: tab1.tex
\begin{deluxetable*}{lccccrrr}
\tabletypesize{\scriptsize}
\tablecaption{Proper Motions for IRS 16SW Co-Moving Group}
\tablewidth{0pt}

\tablehead{
  \colhead{Name} &
  \colhead{K} &
  \colhead{$r_{2D}$} &
  \colhead{$\Delta RA$ (1999.56)} &
  \colhead{$\Delta DEC$ (1999.56)} &
  \colhead{$V_{ra}$}&
  \colhead{$V_{dec}$} &
  \colhead{$V_{radial}$ \tablenotemark{b}} \\
  \colhead{} &
  \colhead{(mag)} &
  \colhead{(arcsec)} &
  \colhead{(arcsec)} &
  \colhead{(arcsec)} &
  \colhead{(km/s)} &
  \colhead{(km/s)} &
  \colhead{(km/s)}
}
\startdata
        16SW & 10.0 &  1.406 &  1.035 $\pm$ 0.008 & -0.951 $\pm$ 0.009 &  228 $\pm$  11 &   61 $\pm$  11 &  400 $\pm$ 40 \\
      16SW-E & 11.0 &  2.153 &  1.836 $\pm$ 0.009 & -1.126 $\pm$ 0.015 &  144 $\pm$  12 &   76 $\pm$  13 &  450 $\pm$ 70 \\
        S2-6 & 12.1 &  2.063 &  1.581 $\pm$ 0.011 & -1.325 $\pm$ 0.013 &  213 $\pm$  12 &   27 $\pm$  13 &  -  \\
        S2-4 & 12.3 &  2.047 &  1.438 $\pm$ 0.012 & -1.457 $\pm$ 0.012 &  207 $\pm$  12 &   60 $\pm$  12 &  -  \\
        S2-5 & 13.3 &  2.051 &  1.883 $\pm$ 0.007 & -0.813 $\pm$ 0.015 &  153 $\pm$  12 &  137 $\pm$  13 &  -  \\
\enddata
\label{tab:members}
\tablenotetext{a}{Positions are with respect to Sgr A* using the absolute
astrometry described in Ghez et al. (2005). All uncertainties 
include absolute astrometric errors.}
\tablenotetext{b}{Radial velocities obtained from Ott Thesis (2003).}
\end{deluxetable*}